# How to Implement Marketing 2.0 Successfully

*Aldhaheri, Abdulrahman[1], Christian Bach[2]*

**Abstract**- The purpose of this research is to develop a model that would close the gap between marketing plans and strategies from one side and the advanced online collaboration applications platforms known as WEB 2.0 in order to implement marketing 2.0 smoothly without disrupting the working environment. We started by examining published articles related to marketing, Web 2.0, Customer Relationship Management Systems, CRM, and social media in a step to conduct an extensive review of the available literature. Then, we presented critique of the articles we have examined. After that, we've been able to develop the model we are proposing in this research. As this paper shows, the proposed model will help in transforming marketing plans and strategies from its traditional approach into, what we would like to call, marketing 2.0 approach smoothly. There are some unavoidable limitations due to the given time and scope constrains. The factors included in the proposed model doesn't cover every related aspect, however, they cover the most important ones.

## INTRODUCTION

The number of the Internet users is growing rapidly. According to a report published by the ICT Data and Statistics Division of the International Telecommunications Union, there are 2.8 Billion users accessing the Internet today, which account to almost 40% of the whole world population [1]. This explosion in the number of the Internet users was due to infrastructural factors, such as the introduction of broadband technologies and their wide availability and, somehow, affordability [2].

The extremely large number of Internet users has driven the technology and fueled the innovations in a sever competition between application developers [2]. One major trend that attracted Internet users was and still is the communication and collaboration applications. Application developers acknowledge the increasing demand of this type of applications by flooding the market with a wide range of communication and collaborations applications, which is now called Web 2.0, or sometimes 'Social Media'. Nowadays, there is a large collection of social media applications available for the Internet users [3-5].

The availability of unbounded communication channels between the Internet users where that can share their experiences, both positive and negative, with each other in a group environment has alarm marketers and business strategists. The Internet users, who are consumers to a certain business, tend to share their experiences about businesses they have dealt with, products the have purchased, or services they have had requested. A game changer; apparently there's a new communication channel marketers can capitalized on [4, 6-8].

At the beginning of the Web 2.0 evolutions, marketers where facing an opportunity that they wanted to take an advantage from. They where concern with the benefits they can achieve [3, 5, 9]. Suddenly, the opportunity turned to be a new valuable channel that has to be incorporated effectively in order to survive in a highly competitive market environment [5, 10-12]. The new rising channel has became a threat to businesses if they reject it, or don't adapt it correctly [13, 14].

[1] University of Bridgeport, School of Engineering, Computer Science and Engineering Department, 126 Park Avenue, Bridgeport, Ct 06604, aaldhahe@my.bridgeport.edu

[2] University of Bridgeport, School of Engineering, Biomedical Engineering Department, 221 University Avenue, room 153, Technology Building, Bridgeport, Ct 06604, cbach@my.bridgeport.edu

This paper discusses the importance on Web 2.0 as a new marketing channel that has a strong influence on consumers and how it should be incorporated into companies strategies as whole, not just the marketing plan. The paper proposes a model that ensures seamless transition and successful implantation of Marketing 2.0 tools, which is the application of Web 2.0 technologies in marketing.

## RESEARCH METHOD

We started by examining published articles related to marketing, Web 2.0, Customer Relationship Management Systems, CRM, and social media in a step to conduct an extensive review of the available literature. Then, we presented critique of the articles we have examined. This has all been done by implementing well-defined scientific research methodologies that were initially proposed and published in [15-19].

Implementing multiple research methods has turned to be critical, such as [18], in particular, for this research, where the scope span two fields and different industries that are marketing and information technology. An extensive review of the literature is important not only for establishing the necessary background, but also as a way to develop a theoretical insight [20].

Another approach that is important in this type of research is to review published case studies. Case studies are full with rich experience based on real world observation [15, 16].

## PROPOSED MODEL

Model Name: Successful Implementation and Adoption of Marketing 2.0

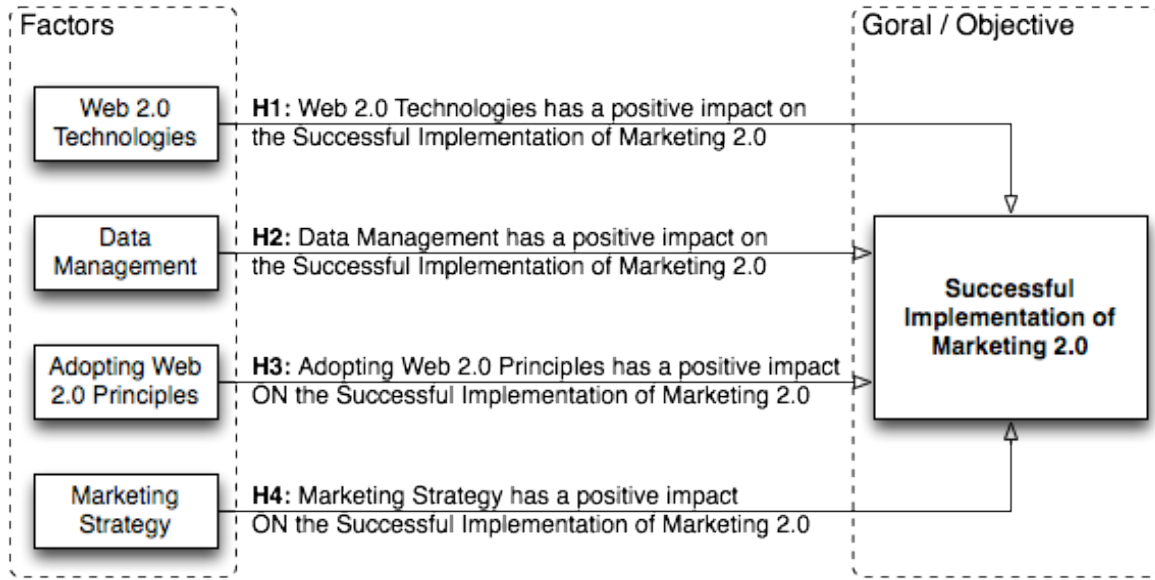

## CHARACTERISTICS OF THE PROPOSED MODEL

The proposed model aims to introduce a clear path through which companies would be able to implement Marketing 2.0 strategies. To be able to do so, first, we'll have to understand what Marketing 2.0 means. This involves defining Web 2.0 and discussing its relation with marketing. Then, we will discuss the factors illustrated in the proposed model by analyzing the hypothesis we've developed on each factor.

## MARKETING 2.0

In order to survive in today's market environment, it has became essential for marketers to reorganize their businesses and make a good use of the Web 2.0 technology and social media [5, 10-12]. The merge between Web 2.0 technology and marketing practice would result in the adaption on what is known as marketing 2.0. Marketing 2.0 is defined as the usage of Web 2.0 technology and the interactive capabilities it provides for the purpose of opening interactive communication channels between consumers and businesses [4].

In order to fully grasp the concept, we need to understand what Web 2.0 means. Web 2.0 is defined as the new generation of websites and web services that capitalize on the collaboration of the users of these services [21, 22]. Taking into confederation how fast and rapid the software industry is moving, it is highly anticipated to see changes in the leadership in Web 2.0 service providers [23]. Therefore, the general definition of Web 2.0 shall be wide enough to cover future services; yet, it would draw a clear line separating Web 1.0 from Web 2.0 technologies.

On the other hand, marketing is defined as the processes used to create, communicate, and deliver value to customers, and to manage customer relationship in a beneficial way for both of the customer and the organization [24]. This actually shows the importance of the communication with consumers for marketers, as well as the importance of the customer's relationship management.

## WEB 2.0 TECHNOLOGIES

**Definitions of Web 2.0**

Web 2.0 technologies have changed the way people interact with each other, which resulted in more interactive communications between consumers. O'Reilly defined Web 2.0 as "The Web as a platform, Harnessing of the Collective Intelligence, Data is the Next Intel Inside, End of the Software Release Cycle, Lightweight Programming Models, Rich User Experiences [25]." Later, he collaborated with Musser and refined the definition as follows: "Web 2.0 is a set of economic, social, and technology trends that collectively form the basis for the next generation of the Internet a more mature, distinctive medium characterized by user participation, openness, and network effects [26]." So, Web 2.0 is the new generation of websites and web services that capitalize on the collaboration of the users of these services [21, 22]. Web 2.0 is a term popularized by O'Reilly Media in media lives international as the mane for service of web development conferences that started in October 2004. After Dale Dougherty mentioned it during a brainstorming session. He suggested that " the web was in a renaissance with changing rules and evolving business models [27]." This idea has turn to be a different type of architecture and different trend in web based application development.

Web 2.0 applications uses a collection of technologies that was developed late in the 90's. These applications allow a large number of users to publish collaboratively, for example, blogs, wikis [28]. The version number in Web 2.0 is commonly used to represent software updates. In Web 2.0 the version number represents an improved form of the World Wide Web. To better understand the concept of Web 2.0 we should discuss the differences between Web 2.0 and it's predecessor, Web 1.0, which is always easier to be done using examples. For example, in web site content management, a web application that is called DoubleClick is a Web 1.0, although it allows you to change and edit the content of your web site [29]. On the contrary, Google AdSense is a Web 2.0 application because it reads your web site and suggest that you add ads on your site that are related to your we site content. Another example that is related to photos and pictures in KodakExpress, which is a Web 1.0 application, allows users to upload pictures to be printed. Flickr, on the other hand, is a Web 2.0 application that allows users to upload and share their photos with its community, which is a community of people sharing photos that could be printed by user.

**Web 2.0 Applications**

The wide range of online applications made it difficult to distinguish between Web 2.0 applications and others that are not Web 2.0 applications. However, the definition of Web 2.0 technologies makes it easier to do so as it encapsulates Web 2.0 applications and web services within the following categories:

1. Bolgs: the word blogs is an abbreviation for Web Logs, which are journals published online by users. Users use different type of contents to enrich their blogs, such as audio and video files. For example, Blogger. Blogs are growing rapidly and gaining popularity faster than other Web 2.0 applications [30].

2. Social Networks: a platform through which users create profiles that could be accessed by other members of the same network. In this model, users can communicate and share content with each other. For example, Facebook. It's work mentioning that the scope and the content of social networks and not usually the same [31].
3. Sharing Communities: applications, or web sites, that allow users to share a specific type of content [4]. For example, YouTube allows users to register and create profiles in order to share videos and comments on these shared videos.
4. Forums: web sites that are dedicated for the discussion of a specialized subject. You can find an online forum for every special interest group nowadays, where members exchange and share their ideas with each other.
5. Content Aggregators: web sites the allow users to control the look and the content for each one by personalizing their portal. This has been made possible through RSS technology, which stands for Real Simple Syndication or sometimes called Rich Sites Summery.

**Web 2.0 Technologies**

One of the most important factors affecting the spread of Web 2.0 technologies is the easiness in their applications development and maintainability. Thanks to advancements in development tools, users with little or non-application development background can now participate in Web 2.0 service development. This has been made possible by the development of tools that are intuitive and easy to use. Some of the technologies that made this possible are:

1. Wikis: web sites that allow users to publish collaboratively. For example, Wikipedia is an online encyclopedia that allows users to create and maintain content.
2. Rich Site Summery, RSS: and sometime Real Simple Syndicate. It allows users to syndicate and customize online contents.
3. Asynchronous JavaScript And XML, AJAX: used in creating dynamic web sites that interact with users inputs and generate content accordingly.

**Application Selection**

Having a wide range of alternatives confuses decision makers when they are selecting a Web 2.0 application to invest in. It is very critical to select the technology that maximizes the business value the most. The chosen application should be able to close gaps in the customer experience and facilitate sophisticated customer segmentation based on click stream transparency and rich web analytics [32]. This could be achieved by selecting the most valuable application for the customers that represent the market the company is targeting.

## DATA MANAGEMENT

**The Challenge**

The amount of data a business expect to receive in case it chooses to facilitate Web 2.0 as channels to communicate with customers is overwhelming. Each day, we create as much as 2.5 Quintillion bytes of data [33]. This is translated as 1 followed by 18 zeros. Of course, not all of them are related to Web 2.0, but it still a lot. For example, there are 12 terabyte of new tweets daily according to IBM [34]. In addition to that, the retrieved information is not in a structured at all. It comes in different format and it's the business responsibility to make sense out of it [35].

**The Solution**

In order to manage information related to customers, a company should have a data management plan that includes having Customer Relationship Management, CRM, capabilities [36]. In order for this plan to succeed [4] it should have the following specifications and capabilities:

- Data Collection
- Data Monitoring
- Data Analysis
- Data Filtering
- Data Classification

- Data Indexing
- Data Retrieval
- Data Interpretation

As Biesdorf et al put it "When a plan is in place, execution becomes easier: integrating data, initiating pilot projects, and creating new tools and training efforts occur in the context of a clear vision for driving business value a vision that's unlikely to run into funding problems or organizational opposition. Over time, of course, the initial plan will get adjusted. Indeed, one key benefit of big data and analytics is that you can learn things about your business that you simply could not see before [37]."

**The Result**

Being able to interact with the customers in two ways communication channels has many advantages. Although it introduces many challenges such as the ones mentioned above. Most of these challenges could be resolved by implementing a data management plan that includes the implementation of a CRM system [38]. Furthermore, such a plan requires long-term commitment and large initial investment.

## ADOPTING WEB 2.0 PRINCIPLES

We've discussed Web 2.0 technologies and the benefits expected of applying these technology to marketing earlier in this paper. However, implementing the technologies is not enough. A more important stage in moving toward Marketing 2.0 is adopting the principles of Web 2.0 [39]. The new approach empower the customers, this fact worries many managers and executive decision makers. According research, "Managers thinking about Web 2.0 adoption are mostly afraid of losing control over communication with clients [21]." The most important thing for most of the managers is the control issue according to the same research.

"Internet users are ready for Web 2.0 concept much better than companies who think about taking advantage of the concept. Organizational culture, risk of losing control over content and simultaneously increasing the power of customers make companies reluctant to the new tools, although decision makers are aware of potential Web 2.0 benefits. The disadvantages of losing control and increasing the power of customers in fact happens no matter whether companies directly would benefit from Web 2.0. Intermediary sites, portals and independent blogs are just a few examples where the information about company, coming from its clients may appear and be popularized. Companies gradually try to accustom to new conditions and leverage the new trends, however the research shows it is very hard what is underlined by saying that "as the three layers of the new media communicative ecology – the social, content and technology – are co-evolving, the Web 2.0 is not only a valuable conceptual aid, but also an important practical imperative touching not only marketing but the whole organization" [39]. [21]."

A company may choose to use more than one Web 2.0 application to implement Marketing 2.0 strategy, it's actually encouraged. However, another important issue companies should focus on when implementing Marketing 2.0 strategy is to synchronize their marketing efforts among the different Web 2.0 applications the company chose to use. Implementing Marketing 2.0 strategy would have negative impact on the companies adopting it if their marketing campaigns appeared to be out of sync and are not timed correctly.

Marketing 2.0 is a two ways communication channel [40]. Therefore, an effective communication strategy is critical for the success of Marketing 2.0 implementation [39]. Having such a strategy will allow the company to maintain and manage its current customers. Meanwhile, it also allows the company to expand its customer base by creating new customers. What seems to scare marketers the most is that they would have to allow customers to communicate to each other if they choose to adopt theses technologies.

## MARKETING STRATEGY

Having the technological infrastructure, applications and data repository, implemented correctly, and adopting Web 2.0 principles completely, wouldn't result in a successful implementation of Marketing 2.0 platform unless the marketing strategy support it completely. Marketing is defined as the processes used to create, communicate, and deliver value to customers, and to manage customer relationship in a beneficial way for both of the customer and the organization [24]. This actually shows the importance of the communication with consumers for marketers, as well as the importance of the customer relationship management. Understanding how customers make their decisions is a key to a successful marketing strategy [41].

The consumer decision journey is a framework developed by Court et al. to identify the point of time at which interaction with customers would result in influencing their decision [41]. The Consumer decision journey spawn through four phases:"

1. Consumer considers an initial set of brands, based on brand perceptions and exposure to recent touch points.
2. Consumers add or subtract brands as they evaluate what they want.
3. Ultimately, the consumer selects a brand at the moment of purchase.
4. After purchasing a product or service, the consumer builds expectations based on experiences to inform the next decision journey [41]."

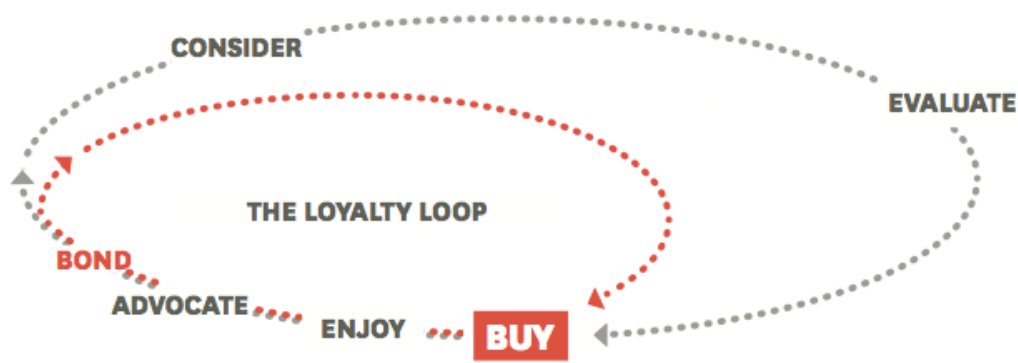

The Consumer Decision Journey [42]

This model provides a clear, realistic understanding of what's really going on. The customer is going through a process when trying to purchase a product. And eventually, turn into a loyal customer that would promote and advocate to the products or the brands they liked. They can do that by sharing their experience online with other potential customers.

Taking this into consideration shows the importance of building new marketing strategy that capitalize on Web 2.0 technologies with the target of touching customers at the points of time that has the most influence on their purchase decision. The proposed Marketing 2.0 strategy should incorporate all of the technological infrastructure and Web 2.0 readiness, data repository and management plan, Web 2.0 principles, and a new marketing strategy that capitalizes on the new capabilities.

## CONCLUSION

In conclusion, Web 2.0 technology has changed the way people interact with each other, which resulted in more interactive communications between consumers. As a result of that, it has became critical for marketers to capitalize on the new communication channels and adapt to Marketing 2.0 as a technology and to update their strategy accordingly. The proposed model will provide an efficient smooth transition for marketers to do so. This include choosing the proper Web 2.0 platform to implement and adapt its principles, manage data efficiently, and update their marketing strategy and plans accordingly. Only by doing this, marketers can be confident that they can influence their customers at the most important touch point throughout the consumer decision journey, which would result in creating a loyal customer.

It is very critical to select the technology that maximizes the business value the most. The chosen application should be able to close gaps in the customer experience and facilitate sophisticated customer segmentation based on click stream transparency and rich web analytics. This could be achieved by selecting the most valuable application for the customers that represent the market the company is targeting. Being able to interact with the customers in two ways communication channels has many advantages. Although it introduces many challenges such as the ones mentioned above. Most of these challenges could be resolved by implementing a data management plan that includes the implementation of a CRM system.

In addition, Marketing 2.0 is a two way communication channel and, therefore, an effective communication strategy is critical for the success of Marketing 2.0 implementation The proposed Marketing 2.0 strategy should incorporate all of the technological infrastructure and Web 2.0 readiness, data repository and management plan, Web 2.0 principles, and a new marketing strategy that capitalizes on the new capabilities.

**Abdulrahman Aldhaheri**

Abdulrahman Aldhaheri is a Ph.D. student at the university of Bridgeport. He holds a Master's degree in Computer Science and a Bachelor degree in Management Information Systems. His research interests include technology management, high performance computing, parallel computing, optimization algorithms, and big data.

**Christian Bach**

Christian Bach is an assistant Professor of Technology Management and Biomedical Engineering at Bridgeport University. He holds a Ph.D. degree in Information Science as well as an executive MBA degree from the University of Albany/SUNY. His multidiscipline research approach is aimed to integrate marketing research and information effectiveness. Some of his research interests include Intracellular Immunization, induced Pluripotent Stem (iPS) cells, Artificial Transcription Factors, Target Detection Assay, Microarrays, Bioreactors, Protein Folding (micro -level), Target Binding Site Computation, micro Database Systems, Knowledge Cubes, Knowledge Management Systems, Collaborative Networks, Global Research Integration, Technology for Advancing Clinical Application.